\documentclass[11pt, onecolumn, twoside, a4paper]{article}
\usepackage{iiai-j}
\usepackage{graphicx}
\usepackage{amsmath}
\usepackage{amssymb}
\usepackage{amsthm}
\usepackage{booktabs}
\usepackage{lipsum}
\usepackage{mathptmx}
\usepackage[hyphens]{url}
\usepackage{comment}
\usepackage[table]{xcolor}

\usepackage{pifont}
\newcommand{\xmark}{\ding{55}}%

\usepackage[rm,up,sc,topmarks,calcwidth,pagestyles]{titlesec}
\titleformat{\section}{\Large\bfseries\rmfamily}{\thesection}{1em}{}
\titleformat{\subsection}{\large\bfseries\rmfamily}{\thesubsection}{1em}{}
\titleformat{\subsubsection}{\large\it\rmfamily}{\thesubsubsection}{1em}{}

\authorshead{M.~Mita, K.~Ito, S.~Ohsawa, H.~Tanaka}
\titlehead{IIAI Journal {\LaTeX} Template}
\vol{1}
\no{1}
\year{2015}
\page{48}
\lastpage{63} 

\title{What is Stablecoin?: A Survey on Its Mechanism and \\Potential as Decentralized Payment Systems}
\author{\hspace{-55pt} Makiko Mita, Kensuke Ito \thanks{The Graduate School of Interdisciplinary Information Studies, The University of Tokyo, Japan}, \\Shohei Ohsawa \thanks{Daisy, Inc., Tokyo, Japan}, Hideyuki Tanaka \thanks{The Graduate School of Interdisciplinary Information Studies, The University of Tokyo, Japan}}
\date{}
\begin{document}
\maketitle
\thispagestyle{empty}

\section*{Abstract}

Our study provides a survey on how existing {\em stablecoins}\textemdash cryptocurrencies aiming at price stabilization\textemdash peg their value to other assets, from the perspective of {\em Decentralized Payment Systems} (DPSs).
This attempt is important because there has been no preceding surveys focusing on the stablecoin as DPSs, i.e., the one aiming at not only price stabilization but also decentralization.
Specifically, we first classified existing stablecoins into four types according to their collaterals (fiat, commodity, crypto, and non-collateralized) and pointed out the high potential of non-collateralized stablecoins as DPSs;
then, we further classified existing non-collateralized stablecoins into two types according to their intervention layers (protocol, application) and confirmed details of their representative mechanisms.
Utilizing concepts such as {\em Quantity Theory of Money (QTM)}, {\em Tobin tax}, and {\em speculative attack}, our survey revealed the status quo where, despite the high potential of non-collateralized stablecoins, they have no standard mechanism to achieve the stablecoin for practical DPSs.
\\
%
%

\noindent
{\it Keywords: }cryptocurrency, decentralized payment system, stablecoin, survey paper

\section{Introduction}

Since Nakamoto \cite{nakamoto2008bitcoin} first proposed their theoretical concept, a large variety of {\em cryptocurrencies}\footnote{The term cryptocurrency has a number of definitions, such as “any form of currency that only exists digitally, that usually has no central issuing or regulating authority but instead uses a decentralized system to record transactions and manage the issuance of new units, and that relies on cryptography to prevent counterfeiting and fraudulent transactions” \cite{dictionary2018definition} and “A medium of exchange that functions like money (in that it can be exchanged for goods and services) but, unlike traditional currency, is untethered to, and independent from, national borders, central banks, sovereigns, or fiats” \cite{maese2016cryptocurrency}. See Houben and Snyers \cite{houben2018cryptocurrencies} for other legal definitions, provided by policy makers.} have been issued and actively traded online. 
After hitting a high of almost \$20,000 for one bitcoin in December 2017 \cite{coinmbit}
, the total market capitalization of cryptocurrencies reached \$796 billion \cite{coinmglo}
\textemdash equivalent to second place in the world ranking of companies by market capitalization at that time, right after Apple Inc.’s \$911 billion \cite{banks}.

From the perspective of online payment systems, several studies \cite{giulia2019decentralized}\cite{kaushal2016bitcoin} have focused on the potential of cryptocurrencies as {\em Decentralized Payment Systems} (DPSs) that can provide various advantages, such as “(i) the diffusion of control among stakeholders; (ii) the ability to engage in trusted commerce without a centralized intermediary; (iii) the potential to disrupt the rents extracted by centralized intermediaries facilitating commerce; and (iv) global consistency and transparency on a shared ledger” \cite{giulia2019decentralized}\footnote{DPSs develop a variety of protocols to address more specific topics, such as micro payments \cite{xu2016blockchain}\cite{10.1145/2810103.2813713} and transparent monetary policy \cite{danezis2016centrally}.}. 
Despite these advantages, however, cryptocurrencies are now difficult to work as practical DPSs due to their {\em high price volatility}\footnote{For specific data on the high price volatility, see, for example, Cryptocurrency Index 30 \cite{cci30}.}.
The high price volatility undermines the three functions of money (i.e., Medium-of-Exchange, Store-of-Value, Unit-of-Account), which results in less demand to own cryptocurrency for online payments.
In fact, according to an online survey in 2018 \cite{dalia}, the awareness of cryptocurrency is 74\% on average in the eight largest cryptocurrency markets (US, UK, Germany, Brazil, Japan, South Korea, China and India), while its ownership remains 7\% on average. 

{\em Stablecoin} is an approach to address this problem of high price volatility, which is for example defined as 
“a digital currency that is pegged to another stable asset like gold, or to major fiat currencies like Euros, Pounds or the US dollar” \cite{stb}\footnote{Note that there are many other definitions for stablecoin, such as “cryptocurrency that has price stable characteristics” \cite{stbah} and “a digital token that will have low price volatility as a result of being pegged to some underlying fiat currency, thereby acting as a store of value, a medium of exchange and unit of accounting for blockchain payments” \cite{hassani2018banking}.}.
Reflecting the necessity for cryptocurrencies as practical DPSs, the market for stablecoin continues to grow rapidly, more than doubling from \$1.4 billion to \$3 billion between 2018 and 2019 \cite{hileman2019state}.

\begin{figure}[t]
 \centering
 \includegraphics[width=0.9\hsize]{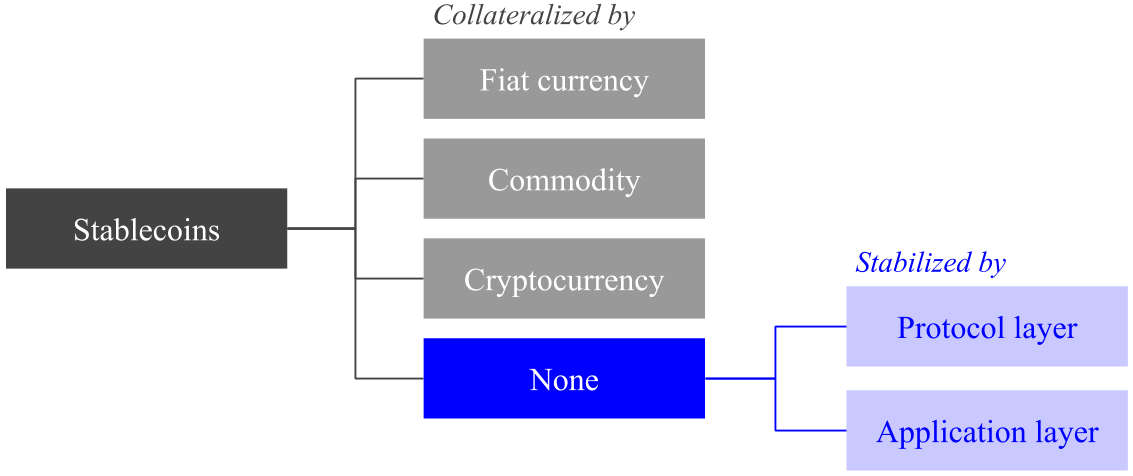}
 \caption{A Classification Tree for Stablecoins}
 \label{tree}
\end{figure}

Our study aims to provide a survey on how such stablecoins peg their value to other assets, from the perspective of DPSs.
This attempt is important because, to the best of our knowledge, there has been no preceding surveys that focuses on the stablecoin as DPSs, i.e., the one aiming at not only price stabilization but also decentralization. 
Specifically, as Figure \ref{tree} depicts, we first classify existing stablecoins into four types according to their collaterals (fiat, commodity, crypto, and non-collateralized)\footnote{This classification is based on the two preceding studies: Zhang et al. \cite{zhang2019regulation} and Mancini-Griffoli \cite{griffoli2018casting}.} and point out the high potential of non-collateralized stablecoins as DPSs; 
then, we further classify existing non-collateralized stablecoins into two types according to their intervention layers (protocol, application) and confirm details of their representative mechanisms. 
In addition, to evaluate a variety of stablecoins, the above process leverages three concepts in economics\textemdash {\em Quantity Theory of Money (QTM)}, {\em Tobin tax}, and {\em speculative attack}.
Our survey consequently highlights the status quo where, despite the high potential of non-collateralized stablecoins, they have no standard mechanism to achieve both price stabilization and decentralization.   
 
The remaining part of this paper is organized as follows. 
Section 2 covers preliminaries, which includes the introduction of the three useful concepts in economics and the four collateral types to classify existing stablecoins.
Section 3 provides an in-depth survey on representative mechanisms in the non-collateralized stablecoin that would have the highest potential as DPSs, by using the layer-based classification.
Finally, Section 4 concludes our survey with describing implications and our next steps.

\section{Preliminaries}

In this section, we first introduce three concepts in economics (QTM, Tobin tax, and speculative attack) which are all useful when considering price-stabilization mechanisms employed in stablecoins.
Moreover, we here classify existing stablecoins into four types according to their collateral (fiat, commodity, crypto, and non-collateralized) and point out that the non-collateralized is the best way to implement stablecoins as DPSs. 

\subsection{Quantity Theory of Money}

QTM is a theory of economics that attempts to explain the price level in terms of the amount of money in circulation.
A representative model in QTM is the {\em equation of exchange} which was formulated by Fisher \cite{fisher1911equation}\cite{fisher2006purchasing} as follows:

\begin{equation}
    M \cdot V = P \cdot Q,
\end{equation}

\noindent
where $M$ is the amount of money supply (in a given period), and $V$ is the velocity of money; $P$ is the price level, and $Q$ is an index of real expenditures on newly produced goods and services.
Namely, the left-hand side represents the scale of an economy through money in circulation, while the right-hand side represents it through goods and services in circulation.
An implication of this equation is that we can adjust the price level $P$ with $M$ and $V$ if their change has no (or small) effect on $Q$ (i.e., if money is neutral). 

From the viewpoint of stablecoins, the QTM is important because it has been a basis of the mechanisms for non-collateralized stablecoins.
As we will confirm in Section 3, all non-collateralized stablecoins 
attempt to stabilize its $P$ by using some mechanism that automatically adjusts $M$ or $V$ in a decentralized manner\footnote{Note that this $P$ denotes not the price of the stablecoin measured by other currency (e.g., USD) but the price of goods and services measured by the stablecoin. Accordingly, the higher $P$ means the lower value of the stablecoin (i.e., inflation).}. 
Specifically, just as central banks tighten monetary policy against inflation, the mechanism decreases $M$ or $V$ if $P$ becomes too high, and vice versa.
Such mechanisms for automatic price adjustment have become more feasible thanks to cryptocurrencies in which we can easily manage transaction records and money supply; on the other hand, there are several related studies preceding cryptocurrencies, including the Tobin tax below.

\subsection{Tobin Tax}

Tobin tax is a proposal by Tobin \cite{tobin1974new}\cite{tobin1978proposal} in 1972, which aims to stabilize the currency exchange rate by penalizing short-term speculative (noise) trading\footnote{Tobin first proposed this idea at a lecture held in 1972, while the proceedings of which \cite{tobin1974new} were compiled in later years.}.
Specifically, it is a small amount (e.g., 0.5\%) of tax imposed on international financial transactions to “throw some sand in the wheels of our excessively efficient international money markets” \cite{tobin1978proposal}\footnote{The Tobin tax and its extensions are often referred to as {\em Currency Transaction Taxes} (CTTs) or {\em Financial Transaction Taxes} (FTTs).}.

A variety of studies have analyzed the Tobin tax.
For example, McCulloch and Pacillo \cite{mcculloch2011tobin} surveyed related literature and available empirical evidence, then concluded that the Tobin tax can be a major source of revenue without causing market distortions while it may not contribute to the stabilization of exchange rate.
To cover its weak contribution to the stabilization, Spahn \cite{spahn1995international} extended the Tobin tax to the two-tier rate structure (Spahn tax) that imposes a higher tax when financial transactions are taken place outside a predetermined range of exchange rate.
Furthermore, Liuzzi et al. \cite{liuzzi2017optimality} leveraged an artificial market to find the optimal rate of the Tobin tax.
Their simulation derived the result close to Spahn \cite{spahn1995international}: a non-negligible level of taxation for highly liquid markets and low (close to zero) levels of taxation for low liquidity markets. 

From the viewpoint of stablecoins, the Tobin tax is important because it has been one of the common tools for existing non-collateralized stablecoins.
As we will confirm in Section 3, several non-collateralized stablecoins adopt the mechanisms which adjust $M$ and $V$ by imposing some amount of fee on each transaction.
Namely, non-collateralized stablecoins inherit the concept of Tobin tax\textemdash “throw some sand in the wheels of our excessively efficient international money markets” \cite{tobin1978proposal}.
In addition, although there are few preceding studies, the Tobin tax should also have implications for collateralized stablecoins, as they have a risk of speculative attack which we will discuss below.

\subsection{Speculative Attack}

Speculative attack is an action that inactive speculators suddenly sell a large amount of currency to deplete the government's reserves, thereby making the pegged exchange rate fail.
Speculators have an incentive to do this attack because they can make big profits from short selling, etc. when the (previously maintained) pegged exchange rate fails.

A variety of studies have analyzed how speculative attacks occur.
For example, Krugman \cite{krugman1979model} was the first study to model the speculative attack targeting a government with insufficient reserves in the foreign exchange market, by extending Salant and Henderson \cite{salant1978market}.
On the other hand, subsequent studies, such as Obstfeld \cite{obstfeld1984rational}\cite{obstfeld1996models} and Chang and Velasco \cite{chang1998financial}, pointed out that speculative attacks (resulting in currency crisis) may occur even in a government with sufficient reserves, by taking into account {\em self-fulfilling features} in which a speculator actually sells the currency if she predicts that other speculators may sell the currency\footnote{See also Diamond and Divig \cite{diamond1983bank} which is the first study to model self-fulfillment in bank runs.}.

From the viewpoint of stablecoins, the speculative attack is important because it has been a main risk for existing stablecoins.
Collateralized stablecoins would likely to be subject to speculative attacks, as they have a similar structure to the pegged currencies relying on government reserves.
Despite this potential risk, to the best of our knowledge, Routledge and Zetlin-Jones \cite{routledge2018currency} is the only preceding study that addresses the speculative attack in collateralized stablecoins. 
Non-collateralized stablecoins would also be subject to the attack, as speculators can profit from their high price volatility or peg failures in a similar way\footnote{Even the Bitcoin protocol has a similar risk, called {\em goldfinger attack} \cite{kroll2013economics} in which miners, even though they have received bitcoin as a reward, damage its value in order to make profits from short selling or holding alternative assets.}.
Thus, as long as stablecoins are (by definition) pegged to other stable assets, we must prevent their mechanisms from speculative attacks.

\subsection{Fiat-, Commodity-, Crypto-, and Non-Collateralized Stablecoins} 


Based on preceding studies \cite{zhang2019regulation}\cite{griffoli2018casting}, this section classifies existing stablecoins into four types according to their collaterals (Figure \ref{tree}).
In addition, for each type, we briefly explain its mechanism and problem as DPSs, which are summarized in Table \ref{table:collateralized}.

The first type is a {\em fiat-collateralized stablecoin} which uses fiat money (e.g., the US dollar) as collateral.
Specifically, it employs a simple and intuitive mechanism that issues new stablecoin on condition that the asset to be pegged is collateralized and, like the gold standard, commits to exchange the stablecoin for collateral at a fixed rate at any time.
A representative example of this type is {\em Tether} \cite{tether}\textemdash a stablecoin which is promised a one-to-one exchange with the US dollar.
Despite this simplicity, however, the fiat-collateralized stablecoin has a problem of requiring a centralized custodian to manage deposited collateral and issue new stablecoins.
This is contrary to the aforementioned advantage of DPSs: “the ability to engage in trusted commerce without a centralized intermediary" \cite{giulia2019decentralized}. 
Accordingly, we cannot use the fiat-collateralized stablecoin as DPSs due to the lack of decentralization.

The second type is a {\em commodity-collateralized stablecoin} which uses commodity (e.g., gold, oil) as collateral.
Although this type uses different collaterals, it actually employs the same mechanism as the fiat-collateralized stablecoin and thus shares the problem, too.
Representative examples of this type are {\em DigixDAO} \cite{digix} and {\em Petro} \cite{petro}\textemdash the stablecoins which are pegged to (and collateralized by) gold in the former and oil in the latter.
Needless to say, we cannot use the commodity-collateralized stablecoin as DPSs due to the same problem: the lack of decentralization.

The third type is a {\em crypto-collateralized stablecoin} which uses cryptocurrency (e.g., bitcoin) as collateral. 
To address the lack of decentralization, it employs a mechanism that uses cryptocurrency (with a decentralized consensus algorithm) for pegging the stablecoin to another stable asset (e.g., the US dollar), which allows any anonymous participants to become a custodian\footnote{Recently, crypto-collateralized stablecoin develops another category referred to as {\em multi-collateralized stablecoin} which uses a variety of assets simultaneously as collateral (e.g., using gold, bitcoin, and Japanese yen for pegging the stablecoin to the US dollar). For example, see {\em Libra} \cite{libra1}\cite{amsden2019libra} and {\em Synthetix} \cite{brookshavven}\cite{synthetix}}. 
A representative example of this type is {\em Dai stablecoin} \cite{daistable}\textemdash a stablecoin pegged to the US dollar but collateralized by the cryptocurrency {\em Ethereum} \cite{wood2014ethereum}\cite{buterin2014ethereum}.
On the other hand, the crypto-collateralized stablecoin has another problem that the mechanism, consisting of at least three assets (stablecoin, cryptocurrency as collateral, the asset to be pegged), is inevitably complicated.
In particular, to prevent the speculative attack while using cryptocurrencies with high price volatility, the mechanism needs much greater collaterals than the value of newly issued stablecoin, which is a well-known {\em over-collateralized problem}\footnote{For example, the issuance of new Dai stablecoin requires at least 150\% collateral (by default).}.
Thus, we cannot use the crypto-collateralized stablecoin as DPSs unless we develop some simpler mechanism without the over-collateralized problem\footnote{Although it is not the main scope of this paper, crypto-collateralized stablecoins have recently developed new mechanisms to avoid the over-collateralized problem, such as {\em Lien protocol} \cite{lienpro}.}.

The fourth type is a {\em non-collateralized stablecoin} which does not use anything as collateral.  
As mentioned above, it employs the mechanism that aims to stabilize $P$ by automatically adjusting $M$ and $V$ in a decentralized manner.
If such a simple and decentralized mechanism were feasible, the non-collateralized stablecoin could obtain the highest potential as DPSs among the four types. 
However, can we really design the mechanism of automatic adjustment, even under a variety of risk including the speculative attack? 
To confirm the feasibility, we in Section 3 provide an in-depth survey on non-collateralized stablecoins, which includes the introduction of representative examples \cite{tiutiun2018usdx}\cite{al2017basis}.
\\

\begin{table}[t]
 \caption{Characteristics of the Four Collateral Types}
 \label{table:collateralized}
 \begin{center}
 \rowcolors{1}{white}{gray!10}
  \begin{tabular}{lcc}
   \toprule
   \textbf{Collateralized by} & \textbf{Decentralization} & \begin{tabular}{c}\textbf{Simplicity}\end{tabular}  \\[2pt]
   \midrule
Fiat currency \cite{tether} &  & \xmark\\[2pt]
Commodity \cite{digix}\cite{petro} &  & \xmark \\[2pt]
Cryptocurrency \cite{daistable} & \xmark & \\[2pt]
None \cite{tiutiun2018usdx}\cite{al2017basis} & \xmark & \xmark\\[2pt]
   \bottomrule
  \end{tabular}
 \end{center}
\end{table}

In this section, we have introduced the three concepts and the four collateral types as preliminaries.
To summarize, the former implies the importance of preceding discussions in economics to consider price-stabilization mechanisms for exisitng stablecoins, and the latter implies the high potential of non-collaterized stablecoins as DPSs.
Table \ref{table:collateralized} represents characteristics of the four collateral types, which reflects (i) fiat- and commodity-collateralized stablecoins are not decentralized because they require some centralized custodian to manage collaterals; (ii) crypto-collateralized stablecoin is not simple because it consists of at least three assets (stablecoin, cryptocurrency as collateral, and the asset to be pegged).
Based on this result, Section 3 focuses on the survey of non-collateralized mechanisms.

\section{Non-Collateralized Mechanisms: A Survey}

This section provides an in-depth survey on non-collateralized mechanisms, by using the classification according to their two intervention layers: {\em protocol} and {\em application} (Figure \ref{tree}).
We here assume that the protocol is a layer related to fundamental consensus-algorithms for the blockchain (e.g., {\em Proof-of-Work} algorithm and {\em Nakamoto consensus} in the Bitcoin protocol), while the application is a layer not related to the consensus-algorithm but for various systems running on the protocol\footnote{See also the following other definitions: "Protocol layer: consists of the core software building blocks that make up a distributed ledger" \cite{hileman20172017}; "Application layer: consists of all applications that are built on existing distributed ledger networks" \cite{hileman20172017}. "The protocol layer lays the foundational structure of the blockchain. It determines the computing language the blockchain will be coded in and any computational rules that will be used on the blockchain" \cite{oecd}; "The application layer is where networks and protocol are used to build applications that users interact with" \cite{oecd}.}.
Such layer-based classification became popular, especially after the Ethereum\textemdash an alternative protocol subsequent to the Bitcoin protocol\textemdash enabled application development on blockchains\footnote{Applications developed on Ethereum are often referred to as {\em Decentralized Applications (DApps)} \cite{raval2016decentralized}.}. 
To the best of our knowledge, non-collateralized stablecoins need to intervene in either protocol or application layers.

Specifically, for protocol and application layers, we confirm their representative mechanisms from the perspective of both proposal and implementation, where the former is based on academic articles and the latter is based on white papers for some project.

\subsection{Stabilization by Protocol Layer}

The survey for protocol layer first confirms two studies, Saito and Iwamura \cite{saito2019make}\footnote{See also Iwamura et al. \cite{iwamura2014can}.} and Saleh \cite{saleh2019volatility}, which both propose modified consensus-algorithms in order for the price stabilization of cryptocurrencies\footnote{Note that these studies updating the Bitcoin protocol do not apply to stablecoin in the definition by Lund \cite{stb}, as they do not envision pegging their value to another asset. However, they are stablecoins in the definition by Tomaino \cite{stbah} which simply defines stablecoin as “cryptocurrency that has price stable characteristics" \cite{stbah}.}.
Subsequently, as a representative implementation, we also confirm the {\em USDX} project \cite{tiutiun2018usdx} which aims to issue stablecoins pegged with the US dollar.

\subsubsection{Proposal}
Studies on the protocol for price stabilization have mainly focused on the consensus algorithm which underlies blockchain-based systems. 
For example, Saito and Iwamura \cite{saito2019make} proposes three modifications to Proof-of-Work algorithm in the current Bitcoin protocol\footnote{Strictly speaking, while their proposal is based on the Bitcoin protocol, it scopes blockchain-based cryptocurrencies in general.}, aiming for the stable price $P$ through autonomous adjustment of the money supply $M$ and the velocity of money $V$.
The first modification limits the re-adjustment of mining difficulty (i.e., Proof-of-Work targets) only when the block-interval exceeds a certain threshold value\footnote{Regardless of the block-interval, the current Bitcoin protocol re-adjust the mining difficulty for every $2,016$ blocks.}.
The second modification makes the amount of mining rewards variable according to each scale of the aforementioned re-adjustment, instead of the existing halving rule\footnote{Regardless of the scale of re-adjustment, the current Bitcoin protocol halves the amount of its mining reward for every $210,000$ blocks.}. 
These two modifications are intended to adjust the growth rate of $M$ (as mining reward) in line with the fluctuating demand (as hash rate).
The third modification imposes a negative interest rate on all bitcoins, which increases in proportion to the time that elapses from the bitcoin issuance just as assets are depreciated over time. 
This modification not only restrains the ever-increasing $M$ but also intervenes in $V$ because, like the Tobin tax, it collects the negative interest for every transaction\footnote{Unlike transaction fees in the current Bitcoin protocol, all bitcoins collected as negative interests will be burned.}\footnote{Tobin tax and this negative interest are different in that the former fixes its own rate while the latter varies its rate depending on time. The ever-increasing negative interest would also have a positive effect on $V$ because it encourages us to change old coin to new one.}.
Saito and Iwamura \cite{saito2019make} tried to make the Bitcoin protocol a practical DPS by the combination of these three modifications. 
	
While Saito and Iwamura \cite{saito2019make} proposed a modified Proof-of-Work algorithm, Saleh \cite{saleh2019volatility} proposed an alternative algorithm for price stability.
He first analyzed Proof-of-Work by the overlapping-generations model \cite{diamond1965national}\textemdash a framework used in economics\textemdash and pointed out that it can cause exceptional price volatility and welfare impairment.
To solve the problem, Saleh \cite{saleh2019volatility} recommends us to adopt {\em Proof-of-Burn} algorithm\footnote{Note that, prior to Saleh \cite{saleh2019volatility}, Stewart \cite{pob} for the first time proposed the concept of Proof-of-Burn in 2012.} in which, to create new blocks, miners need to send a certain amount of their coins to an unspendable (locked) address.
Here, the Proof-of-Burn adjusts mining difficulty through the amount of coins that should be burned; in other words, his proposal controls $M$ by adjusting the amount of coins burned as mining costs, against the amount of newly issued coins as mining rewards.

One of the problems with these protocols is that their price stabilization does not use market price (e.g., BTC/USD) directly, but uses mining difficulty as a proxy variable.
It is still controversial whether the mining difficulty (or hash rate in the case of Bitcoin protocol) can be a proxy variable for market prices \cite{georgoula2015using}\cite{kristoufek2015main}; however, at least on a yearly basis, there seems to be no explicit correlation between the two \cite{difficulty}.
In any case, these protocols can no longer stabilize prices under a chronic deviation between market prices and mining difficulty (i.e., unstable proxy variable).
Accordingly, in \ref{proimple}, we introduce an implementation of the non-collateralized and protocol-based stablecoin, which aims to peg their price directly with the US dollar.

\subsubsection{Implementation} \label{proimple}
Tiutiun et al. \cite{tiutiun2018usdx} proposed the USDX\textemdash a project for non-collateralized stablecoin based on the protocol layer.  
Specifically, this project has issued a stablecoin called USDY\footnote{This project contains two coins: USDX and USDY, where the former is a type of governance token and the latter is stablecoin pegged to the US dollar.}, which aims to maintain the same price with the US dollar through the following three mechanisms. 
The first mechanism is {\em variable block reward} \cite{tiutiun2018usdx} to adjust the amount of mining rewards, just as the second modification in Saito and Iwamura \cite{saito2019make}.
On the other hand, to adjust mining rewards, the USDX leverages the USDY/USD price index\footnote{Here, we have another important problem: how to obtain data outside the blockchain (e.g., the USDY/USD price index) in a decentralized manner while ensuring their reliability? In the context of blockchain, this is often referred to as the {\em Oracle problem} \cite{voshmgir2019token}. Specifically, although we will spare you the details, the USDX employs an original mechanism called the decentralized Schelling point Oracle system.}, rather than the mining difficulty. 
That is, when USDY/USD $> 1$, the amount of mining rewards (USDY) increases in order for the price reduction (by increasing the growth rate of $M$), and vice versa.
The second mechanism is {\em lock-in mining} \cite{tiutiun2018usdx} which is a kind of emergency measure, activated only when the state of USDY/USD $< 1$ persists even if the variable block reward takes the lowest value.
To further reduce the money supply $M$, the lock-in mining allows users to (stochastically) create a new block by temporary locking their own USDY for a predetermined period which varies according to the rigidity of USDY/USD $< 1$\footnote{Unlike the Proof-of-Burn, the locked USDY will back to the holders after the predetermined period.}.
When users successfully create new blocks with the lock-in mining, they can receive mining rewards (USDY) independent of the variable block reward. 
The third mechanism is {\em variable transaction fee} \cite{tiutiun2018usdx}, which is the same with the third modification in Saito and Iwamura \cite{saito2019make} in that it burns circulating coin with a variable rate to control $M$ and $V$.
On the other hand, the USDX also leverages the USDY/USD price index to determine the rate of transaction fee, rather than the time from the coin issuance.
That is, when USDY/USD $>$ 1, the amount of transaction fee decreases in order for the price reduction (by increasing both $M$ and $V$), and vice versa.
The USDX project aims to peg its stablecoin USDY with the US dollar by the combination of the three mechanisms above.

The problem with the USDX is that its mechanism leads to the unstable {\em purchasing power} (i.e., the amount of USDY $\cdot$ the current price of USDY) in each wallet.
While stablecoin, by definition, focuses on stabilizing its price against other assets, it needs to stabilize the purchasing power of users as well in order to be a practical DPS satisfying the store-of-value.
In the case of USDX, the variable block reward and the lock-in mining intervene only in the purchasing power of miners who voluntarily join the mechanisms; however, the variable transaction fee intervenes in that of individuals who do nothing but hold USDY.
Furthermore, although the Tobin tax and negative interest rate in Saito and Iwamura \cite{saito2019make} do the same type of intervention, the variable transaction fee in USDX would be worse because its fee rate fluctuates unpredictably (according to USDY/USD).
Thus, to enable a constant purchasing power, we should controll $M$ and $V$ through mechanisms based on some voluntary action (e.g., mining), and if it were inevitable to use transaction fees, the rates should at least be predictable.

\subsection{Stabilization by Application Layer}
The survey for application layer, on the other hand, first confirms three studies, Ametrano \cite{ametrano2016hayek}, Morini \cite{morini2014inv}, and Sams \cite{sams2015note}.
Even though Ametrano \cite{ametrano2016hayek} and Morini \cite{morini2014inv} assume direct intervention in each wallet, Sams \cite{sams2015note} inherits these studies and proposes another mechanism\textemdash {\em Seigniorage Share}.
Subsequently, we also confirm the detail of the {\em Basis} project \cite{al2017basis} as a representative implementation of the Seigniorage Share.

\subsubsection{Proposal} 
Discussions on the application for price stabilization stem from Ametrano \cite{ametrano2016hayek}, which proposed a non-collateralized stablecoin named {\em Hayek Money}\footnote{It was named after the denationalization of money \cite{hayek1990denationalisation}\textemdash the concept proposed by Hayek.}.
The gist of the Hayek money is its {\em rebasement} mechanism \cite{ametrano2016hayek} that automatically adjusts $M$ by modifying (not consensus-algorithm but) the amount of money stocked in each wallet, according to the data on current price which is updated whenever miners create new blocks.
However, as with the USDX project, this simple mechanism leads to the unstable purchasing power because the rebasement directly modifies the amount of money in each wallet.

To address this problem, Morini \cite{morini2014inv} argued that the Hayek money should divide its wallet into two types: {\em Inv wallets} for investment and {\em Sav wallets} for saving. 
Here, in order to leave the option of stable purchasing power, the rebasement only intervenes in the Inv wallets (accordingly, $M$ in Inv wallets is subject to the higher fluctuation than the original Hayek money, to cover the adjustment for $M$ in Sav wallets).
Namely, Morini \cite{morini2014inv} allows users to allocate their holding money into both Inv and Sav wallets, thereby offering “the freedom to choose how much they want to be affected by changes of money supply” \cite{morini2014inv} along with their risk appetite. 
This mechanism appears to be effective as it only intervenes in the money whose owners have accepted the unstable purchasing power; however, it has another problem.  
Consider the case where most users predict an increasing trend of $P$ (i.e., a decline in the price of Hayek money).
In this case, decreasing $M$ for price-stabilization becomes difficult because speculators would transfer their money from the Inv wallet to the Sav wallet in order to avoid the rebasement.
To make matters worse, this Inv-to-Sav transfer will impose the higher decreasing rate on the Inv wallets, which further accelerates the Inv-to-Sav transfer.
Therefore, if speculators could freely transfer their money between Inv-Sav wallets (i.e., the Inv/Sav ratio is unstable), it would be difficult to adjust (especially to reduce) $M$ in the Inv wallet\footnote{In response to the problem, Morini \cite{morini2014inv} suggests collecting a small amount of fee from Sav wallets to decrease $M$, only if there are extremely few money in Inv wallets. However, this is contrary to the original purpose of stabilizing the purchasing power in Sav wallets.}.

Inhereting the above discussion, Sams \cite{sams2015note} proposed a new mechanism for the price stabilization in application layer, named Seigniorage Share.
As the name implies, the mechanism aims to automatically adjust $M$ through {\em shares} with which users can purchase stablecoins.
Specifically, when the mechanism detects an increasing trend of $P$\footnote{As with the Hayek money, Sams \cite{sams2015note} assumes that the data on current price is updated whenever miners create new blocks. Moreover, Sams \cite{sams2015note} mentions the possibility of using mining difficulty as a proxy variable, as in models for protocol layers, in addition to using general oracles that obtain price data from outside the system (e.g., exchanges).}, it issues new shares to decrease the current $M$.
Shares are distributed among bidders who burned an arbitrary amount of holding coins in a decentralized auction\footnote{See original article for the detail of decentralized auctions.}.
Conversely, when the mechanism detects a decreasing trend of $P$, it issues new coins to increase the current $M$.
Coins are distributed among bidders who burned an arbitrary amount of holding shares in another decentralized auction. 
In other words, Seigniorage Share is a mechanism that intends for the price stabilization by letting users voluntarily balance the amount of coins and shares, where, unlike the free transfer between Inv-Sav wallets, the coin-share exchange rate is determined in auctions.


\subsubsection{Implementation} 
Al-Naji et al. \cite{al2017basis} proposed the Basis\textemdash a project for non-collateralized stablecoin based on the application layer.
Specifically, this project has issued a stablecoin called Basis token, which aims to maintain the same rate with the US dollar through the Seigniorage Share.
On the other hand, to make the mechanism more practical, the Basis project has made some modifications to the original Seigniorage Share \cite{sams2015note}.
One of the biggest modifications is to adjust $M$ with another token called {\em bonds}.
Bonds are similar to shares in that they are newly issued to reduce $M$ and are distributed in a decentralized auction (where bonds are sold for prices of less than 1 basis); however, they are not used to purchase stablecoins but, as with real bonds, redeemed at a fixed exchange rate of Bond/Basis = 1 when the mechanism newly issues the Basis token in the future.
That is, the mechanism encourages users to purchase the bond by the commitment to redeem it with the newly issued Basis token\footnote{It is confusing but the Basis project also uses shares which has a different role from those in Sams \cite{sams2015note}. Shares in the Basis project are only issued at the genesis of the blockchain, and shareholders can receive the newly issued Basis tokens, when the mechanism needs to increase $M$ even after all bonds have been redeemed.}.
This modification\textemdash using bonds instead of shares\textemdash is primarily for the robustness against the case of high $P$ (i.e., inflation). 
In the original Seigniorage Share \cite{sams2015note}, stablecoin would be difficult to recover from extremely high $P$ because the more new shares are supplied, the lower their price and consequently the less power to reduce $M$. 
The Basis project attempts to maintain this power to reduce $M$, by adopting bonds with the commitment to redeem at a fixed exchange rate\footnote{In addition, the bonds in the Basis project will expire after five years, even if they are not redeemed with the Basis tokens. This would help maintain the price of new bonds in terms of reducing the amount of existing bonds.}.

However, regardless of this modification, the Basis project and the original Seigniorage Share \cite{sams2015note} both have a critical problem which is summarized as follows\textemdash “incentives to buy bonds are based on a circular dependency: people will buy the bonds if they think Basis will climb up, but Basis will only climb up if people buy the bonds” \cite{circular}.
In other words, there is a kind of tautology in the Seigniorage Share that requires a decrease of future $P$ in order to decrease current $P$ (i.e., shares or bonds are in demand).
It is highly doubtful that speculators would buy shares or bonds under such a tautological mechanism.
Perhaps due to this problem, the developer team announced the closure of Basis project in December 2018 \cite{basisclose}, even though it raised \$133 million through {\em Initial Coin Offering} (ICO).
\\

\begin{table}[t]
\caption{Stabilization Mechanisms for Non-Collateralized Stablecoins}
\vspace{10pt}
\label{table:non-col}
\resizebox{\textwidth}{!}{%
\rowcolors{1}{white}{gray!10}
\begin{tabular}{lccccl}
\toprule
 & \multicolumn{2}{c}{\textbf{Intervention Layers}} & \multicolumn{2}{c}{\textbf{QTM}} & \multicolumn{1}{c}{\textbf{Problems}} \\[2pt]
\cmidrule{2-6}
 & Protocol & \multicolumn{1}{l}{Application} & \multicolumn{1}{l}{$M$} & \multicolumn{1}{l}{$V$} & \multicolumn{1}{l}{} \\[2pt]
\midrule
\textbf{Proposals} & \multicolumn{1}{l}{} & \multicolumn{1}{l}{} & \multicolumn{1}{l}{} & \multicolumn{1}{l}{} & \multicolumn{1}{l}{} \\[2pt]
\hspace{10pt}Proof-of-Work \cite{saito2019make} & \xmark &  & \xmark & \xmark & unstable proxy variable \\[2pt]
\hspace{10pt}Proof-of-Burn \cite{saleh2019volatility} & \xmark &  & \xmark &  & unstable proxy variable \\[2pt]
\hspace{10pt}Hayek Money \cite{ametrano2016hayek} &  & \xmark & \xmark &  & unstable purchasing power\\[2pt]
\hspace{10pt}Inv/Sav wallets \cite{morini2014inv} &  & \xmark & \xmark &  & unstable Inv/Sav ratio\\[2pt]
\hspace{10pt}Seigniorage Share \cite{sams2015note} &  & \xmark & \xmark &  & tautological mechanism \\[2pt]
\textbf{Implementations} &  &  &  &  &  \\[2pt]
\hspace{10pt}USDX \cite{tiutiun2018usdx} & \xmark &  & \xmark & \xmark & unstable purchasing power\\[2pt]
\hspace{10pt}Basis \cite{al2017basis} &  & \xmark & \xmark &  & tautological mechanism \\[2pt]
\bottomrule
\end{tabular}%
}
\end{table}

In this section, we have surveyed non-collateralized mechanisms in protocol and application layers, from the perspective of both proposal and implementation.
To summarize, as Table \ref{table:non-col} shows, all existing non-collateralized stablecoins are not practical as DPSs due to some problem. 
In the protocol layer, proposals \cite{saito2019make}\cite{saleh2019volatility} have the problem of unstable proxy variable for market price, and even though an implementation \cite{tiutiun2018usdx} uses price data outside the system, it has another problem of unstable purchasing power (of each wallet).
In the application layer, no mechanism has resolved this unstable purchasing power.
Despite proposals to increase the type of wallets \cite{morini2014inv} and tokens \cite{sams2015note}, speculators would not voluntarily contribute to the price stabilization.  
This is implied by the closure of a representative project \cite{al2017basis} for the application layer\footnote{Despite the closure, non-collateralized stablecoin still has a potential as a DPS and continues to develop new mechanisms such as Terra \cite{kereiakes2019terra}. Re-investigating the ever-increasing new proposals would be one of the future works of this survey.}.

\section{Conclusion}
In this paper, we for the first time surveyed how existing stablecoins\textemdash cryptocurrencies aiming at price stabilization\textemdash peg their value to other assets, from the perspective of DPSs.
Specifically, our survey first classified existing stablecoins into four types according to their collaterals (fiat, commodity, crypto, and non-collateralized) and pointed out the high potential of non-collateralized stablecoins as DPSs (Table \ref{table:collateralized}); then, it further classified existing non-collateralized stablecoins into two types according to their intervention layers (protocol, application) and surveyed proposals and implementations for each layer (Table \ref{table:non-col}). 
This survey focusing on non-collateralized stablecoins pointed out that, due to a variety of problems, all existing mechanisms cannot ensure the constant purchasing power in each wallet which may be owned by speculators.

This result implies the status quo where, despite the high potential of non-collateralized stablecoins, they have no standard mechanism to achieve both price stabilization and decentralization.
Accordingly, in order to make cryptocurrencies practical DPSs, our next step would be to design some new non-collateralized mechanism that enables a constant purchasing power, while taking into account the aforementioned concepts such as QTM, Tobin tax, and speculative attack.

\section*{Acknowledgments}

We would like to express our gratitude to B Cryptos for providing valuable comments and financial support. 
\\
We also would like to express our gratitude to referees and participants in IIAI AAI 2019 for providing valuable comments.

\bibliographystyle{ieeetr}
\bibliography{references}
\end{document}